\documentclass[aps,prb,twocolumn,superscriptaddress,10pt]{revtex4-2}
\usepackage{siunitx}
\usepackage{amsfonts}
\usepackage{amsmath}
\usepackage{amssymb}
\usepackage{graphicx}
\usepackage{dcolumn}
\usepackage{mciteplus}
\usepackage{euscript}
\usepackage{multirow}
\usepackage{color}
\usepackage{bm}
\usepackage{wasysym}  

\graphicspath{{./figs/}}
\def\ms{M_\text{s}}

\newcommand{\bfM}{{\bf M}}
\newcommand{\bfm}{{\bf m}}
\newcommand{\bfH}{{\bf H}}
\newcommand{\bfh}{{\bf h}}

\newcommand{\memome}[1]{}
\newcommand{\req}[1]{Eq.~(\ref{#1})}

\newcommand{\rfig}[1]{Fig.~\ref{#1}}
\newcommand{\rFig}[1]{Figure~\ref{#1}}

\newcommand{\ud}{\mathrm{d}}

\newcommand{\miold}[1]{\iffalse{#1}\fi}

\begin{document}

\title{Phase-Factor-Controlled Surface Spirals in the Magnetic Conical Phase: The Role of In-Plane Directionality}

\author{Haijun Zhao}
\thanks{Correspondence to: haijunzhao@seu.edu.cn}
\affiliation{School of Physics, Southeast University, 211189 Nanjing, China.}
\affiliation{Ames National Laboratory, U.S. Department of Energy, Ames, Iowa 50011, USA}
\author{Tae-Hoon Kim}
\affiliation{Ames National Laboratory, U.S. Department of Energy, Ames, Iowa 50011, USA}

\author{Lin Zhou}

\affiliation{Ames National Laboratory, U.S. Department of Energy, Ames, Iowa 50011, USA}
\affiliation{Department of Materials Science and Engineering, University of Virginia, Charlottesville, VA 22904}
\author{Liqin Ke}
\thanks{Correspondence to: liqin.ke@virginia.edu}
\affiliation{Ames National Laboratory, U.S. Department of Energy, Ames, Iowa 50011, USA}
\affiliation{Department of Materials Science and Engineering, University of Virginia, Charlottesville, VA 22904}

\date{\today}

\begin{abstract}
 In chiral magnets, the magnetic textures surrounding domain walls exhibit a rich variety of structures, offering insights into fundamental physics and potential applications in spintronic devices. Conical spirals and related structures possess intrinsic in-plane directionalities governed by phase factors $\phi_0$, which are often obscured in long spirals due to cylindrical symmetry but become prominent in short spirals or thin films. Using micromagnetic simulations, we systematically studied magnetic textures at ferromagnetic-conical interfaces (FCI), including 1D and 2D FCIs with various shapes. Surface spirals (SS) emerge adjacent to these FCIs, closely linked to the cone's in-plane reorientation. In 1D FCIs, reorientation controls the presence, shape, and topological charge of the SS, with a discontinuity point observed where spirals with opposite charges form on opposite sides. In 2D FCIs, eyebrow-like SS are evident. The reorientation angle between top and bottom SS is controlled by the film thickness, similar to stacked spirals reported previously. We further demonstrate that SSs form at the facets of skyrmion clusters within the conical phase, as confirmed by both simulations and  Lorentz transmission electron microscopy observations in Co$_8$Zn$_{10}$Mn$_2$ thin films. The experiments specifically reveal two distinct formation pathways: thermally activated co-growth and field-driven transformation from residual helices. These findings establish $\phi_0$ as a fundamental control parameter for magnetic states, enabling promising spintronic functionalities such as multi-state memory through SS polymorphism and energy-efficient neuromorphic computing via controlled topological transitions.
\end{abstract}

\maketitle

\section{Introduction}
Chiral magnetism has garnered significant research interest owing to its profound implications in fundamental science and technological advancements \cite{schoenherr2018Topological_,bogdanov2020Emergence_,kanazawa2017Noncentrosymmetric_,fert2017Magnetic_,tokura2021Magnetic_}. In chiral magnetic materials, the intricate interplay between Heisenberg exchange and Dzyaloshinskii-Moriya interaction (DMI), stemming from relativistic spin-orbit coupling and inversion symmetry breaking, generates a rich tapestry of spin configurations \cite{muhlbauer2009skyrmion_,foster2019Twodimensional_,kim2020mechanisms_,yu2010realspace_,ran2021Creation_}. Heisenberg exchange inherently favors collinear spin alignment, whereas DMI induces a subtle twist between adjacent spins, fostering the emergence of spin spirals. The competition leads to the formation of various magnetic textures that are either topologically trivial or nontrivial. 

The topologically trivial structures normally have spiral propagating along one direction and collinear spin alignment in the plane perpendicular to the propagation direction.
An example of such structures is the helical spin spiral state~\cite{dzyaloshinskii1965theory_,milde2013Unwinding_,leonov2016chiral_,rybakov2016new_,schoenherr2018Topological_,kim2025Topological_}. Due to the isotropic nature of exchange and DMI, the helical spiral can orient in any direction, often resulting in a disordered multi-domain state. However, anisotropy introduced by factors such as easy-plane magnetic anisotropy, strain-induced modulation of DMI, or even current-driven orientation preferences, can align these spirals into ordered monodomains. Moreover, an external magnetic field exerts an influence, favoring a spiral direction parallel to its own, causing spins to tilt towards the field's direction to minimize Zeeman energy. This progression culminates in the formation of a conical magnetic state, where the tilt angle gradually increases with the field's intensity until, at saturation, all spins align fully parallel to the external field, transitioning into a ferromagnetic (FM) state~\cite{zheng2018Experimental_,milde2013Unwinding_}. 

The topologically nontrivial magnetic textures are normally merged within a topologically trivial magnetic texture, or forming lattices. The topological protection making them act like quasi-particles causing them potential data carriers in future digital devices \cite{zhou2018Magnetic_}. An important example of such structures is skyrmions~\cite{muhlbauer2009skyrmion_,yu2010realspace_,kim2021Kinetics_,kim2020mechanisms_}, which were extensively studied for their potential in pioneering spintronic technologies, including advanced information storage systems \cite{tomasello2015Strategy_,zheng2018Experimental_}, logical devices \cite{luo2021Skyrmion_}, and unconventional computing paradigms \cite{li2021Magnetic_a}. 

When comparing films with bulk samples, it becomes evident that the surface of films induces additional twists \cite{rybakov2013Threedimensional_,rybakov2016new_,leonov2016chiral_,hals2017New_,meynell2014Surface_,hellman2017Interfaceinduced_}. These twists subsequently modify the magnetic structures in two significant ways. Firstly, they finely tune the complexities of the existing magnetic structures. For instance, both theoretical investigations \cite{rybakov2013Threedimensional_,rybakov2016new_,leonov2016chiral_} and experimental observations \cite{zhang2018direct_} reveal that the helicity of a skyrmion deviates from its midsection value as it approaches the surface, transforming pure Bloch-type skyrmions into partially Neel-type configurations. Secondly, these surface twists can also give rise to novel magnetic structures, such as skyrmion bobbers \cite{rybakov2015New_,zheng2018Experimental_,ran2021Creation_,redies2019Distinct_,kim2020mechanisms_}, stacked spirals~\cite{rybakov2016new_}, and other boundary-driven magnetic twist states \cite{hals2017New_,song2018Quantification_,jin2017Control_,meynell2014Surface_}. Specifically, stacked spirals comprise a conical spiral midsection and surface-induced spirals localized in the superficial layers, exhibiting a finite penetration depth. The stacked spirals display a stripe-like pattern that mimics the helical phase. However, they provide an additional layer of flexibility in their bending direction during penetration, enabling them to undergo full-range reorientation (spanning from 0 to 2$\pi$) rather than the limited half-range reorientation (0 to $\pi$) typical of helical structures. Furthermore, the angle between the propagation directions of the stacked spirals on the top and bottom surfaces changes with the film's thickness.

There exists a further remarkable distinction between films and bulk samples, rooted in the distinct symmetry characteristics of their magnetic structures. This distinction has garnered limited attention thus far. Experimentally observable properties are typically net average values. In bulk samples, long spirals often display cylindrical symmetry, yet they develop a net anisotropy in the plane perpendicular to their propagation direction, leading to an apparent in-plane circular symmetry. However, in thin films or when focusing on a small section of layers such as superficial layers, the inherent non-circular symmetry becomes pronounced due to spiral shortening, imparting a distinct in-plane directionality. Regarding the aforementioned stacked spirals~\cite{rybakov2016new_}, it is reasonable to expect that their optimal propagation direction is determined solely by the directionality of the cone in thick samples (even though the entire cone surface is covered by stacked spirals, one can define this directionality by extending the inner conical phase to the surface). Therefore, the variation in the angle between the propagation directions of stacked spirals on the top and bottom surfaces with thickness can be directly attributed to the relative angle between the directionality of the cone on the two surfaces, which is solely a function of thickness. Another clear instance of this directionality becomes apparent in the interaction between skyrmions embedded within the conical phase. The ``approaching direction" between these skyrmions attains significance, as the reorientation angle—defined as the angle between the conical phase's directionality and the direction of approach—now exerts an influence on their interaction. Consequently, this interaction becomes contingent not merely on distance but also on directionality \cite{kim2020mechanisms_}. A more pronounced exhibition of this directionality, or in-plane anisotropy, is observed in 2D films, where geometry-induced anisotropy emerges due to factors such as a tilted field or tilted magnetic anisotropy \cite{kameda2021Controllable_}.

In principle, controlling this directionality is straightforward. Reference \cite{delser2021Archimedean_} proposes using a perpendicular circularly polarized AC field to screw the conical phase, thereby modifying its directionality. By the same mechanism, magnons—which have been shown to switch the magnetization of a ferromagnetic (FM) cuboid \cite{wang2019Magnetization_}—would also screw the phase, enabling a more integrated design.
Consequently, manipulating this directionality presents a potential avenue for controlling the formation of magnetic textures, which opens promising avenues for designing novel spintronic devices, such as ultrahigh-density memory elements and neuromorphic
emulation, as will be discussed in the context of our findings.

Despite its importance, the in-plane directionality of spirals, particularly in the conical phase, has been understudied. This research delves into this underexplored territory by examining magnetic structures formed at superficial layers near domain walls separating the conical phase from other phases, like FM domains or skyrmion clusters. This choice is motivated by three key factors: the domain wall's normal direction as an in-plane reference, domain walls' ability to stabilize diverse, orientation-sensitive magnetic structures, and the conical phase's capacity to host magnetic objects like bobbers \cite{kim2020mechanisms_} and stacked spirals~\cite{rybakov2016new_}, primarily interacting with the conical background within superficial layers.

The paper is organized as follows: Initially, we undertake a numerical investigation of magnetic structures in the vicinity of ferromagnetic-conical interfaces (FCI). We explore various idealized FCI configurations, including a one-dimensional (1D) flat FCI and two-dimensional (2D) FCIs with square and circular cross-sections. By adjusting the in-plane dimensionality, or equivalently, by altering the orientation of domain walls, we carefully examine how these changes influence the formation and evolution of surface magnetic structures. Subsequently, we shift our focus to a numerical study where we replace the FM domain with skyrmion clusters. We then compare our findings with observations obtained through Lorentz transmission electron microscopy (LTEM) in $\beta$-Mn-type Co$_8$Zn$_{10}$Mn$_2$ thin films. Concluding this progression, we examine potential spintronic applications arising from phase-factor-controlled magnetic phenomena, with specific consideration given to high-density memory implementations and neuromorphic computing paradigms. Finially, we give the conclusions.

\section{Numerical Method}
We utilize the standard micromagnetic model to simulate magnetic configurations, expressed as
\begin{eqnarray}
  E &=& E_{\text{ex}} + E_{\text{DM}} + E_{\text{Zeeman}} \nonumber\\
  &=& A\,(\nabla\mathbf{m})^2 + D\,\mathbf{m}\!\cdot\!(\nabla\times\mathbf{m}) - \mu_0\mathbf{M}\!\cdot\!\mathbf{H}.
  \label{eq:energy}
\end{eqnarray}
Here, $E_{\text{ex}}$, $E_{\text{DM}}$, and $E_{\text{Zeeman}}$ denote exchange, Dzyaloshinskii-Moriya, and Zeeman energies, respectively.  
The magnetization vector is $\mathbf{M}=M_{\text{s}}\mathbf{m}$, with saturation amplitude $M_{\text{s}}$ and unit vector $\mathbf{m}=[m_x,m_y,m_z]$.  
$A$ and $D$ are the exchange stiffness and DM constants; $\mu_0$ is the vacuum permeability; and the external field $\mathbf{H}=[0,0,H]$ is applied along $\hat{z}$.

Our study targets the conical phase, whose stability depends on $H$.  
The conical phase possesses propagation vector along $\hat{z}$ (see the “Cone” region in Fig.~1(a)).  
$L_D=4\pi A/|D|$ and the saturation field $H_D=D^2/(2\mu_0 A \ms)$ are determined by the interplay between exchange and DM interactions \cite{rybakov2016new_,leonov2016chiral_,dzyaloshinskii1965theory_,bogdanov1994Thermodynamically_}.
Inside the cone, collinear layers exhibit helical rotation of the in--plane magnetization along $\hat{z}$ whenever $H\!<\!H_D$.  

The magnetization at height $z$ is \cite{rybakov2016new_,leonov2016chiral_}
\begin{equation}
  \bfm(z)=\bigg[\sqrt{1-m_z^2}\cos\phi(z), \sqrt{1-m_z^2}\sin\phi(z), m_z\bigg],
  \label{eq:m}
\end{equation}
where, $m_z=H/H_D$ is a constant and the phase factor
\begin{equation}
  \phi(z)=\pm2\pi z/L_D+\phi_0
  \label{eq:lz_vs_phi}
\end{equation}
describes the spiraling. The sign of the phase factor dictates the chiral direction (clockwise or counterclockwise), which is governed by the sign of $D$.

In the ideal conical state, the net in-plane magnetization averaged over one pitch vanishes:
\[
\frac{1}{L_D}\int_{z}^{z+L_D}\! m_{x,y}(z')\,dz' = 0,
\]
yielding a predominantly axial ($\parallel\hat{z}$) demagnetizing field. Under strong fields ($H \gtrsim 0.3H_D$), demagnetization corrections are subdominant to Zeeman energy, mainly renormalizing the effective field. However, notable dipolar effects occur in three regimes:

(i) \textit{Low-field metastability} ($H \lesssim 0.3H_D$): The field-aligned single-Q conical phase becomes metastable with reduced energy barrier, allowing nucleation of helical phases with propagation vectors deviating from $\hat{z}$.

(ii) \textit{Ultra-thin film constraint} ($L_z \ll L_D$): Geometrical confinement limits conical phase formation, while demagnetization effects further disrupt the spin texture.

(iii) \textit{Non-periodic thickness} ($L_z \neq nL_D$): Residual in-plane components emerge, but their \textit{areal density} $\propto 1/L_z$ becomes negligible in thick films ($L_z \gg L_D$).

For our parameters ($H \geq 0.4H_D$, $L_z = 8L_D \gg L_D$), conical stability is maintained and demagnetization effects are minimized. Following Refs.~\cite{rybakov2016new_,leonov2016chiral_}, we therefore neglect dipolar interactions, noting that the excluded regimes require dedicated treatment.

For each layer, the conical phase has an in-plane directionality that aligns with $(\cos\phi(z), \sin\phi(z), 0)$ (see Fig. 1(b)). However, to detect this directionality, a reference direction is necessary. In Figure 1, we illustrate the geometry of an FCI, where the normal direction of the domain wall (traversing from the FM domain to the conical domain) serves as the reference direction, designated as the positive $\hat{x}$ axis. Averaging the cone along the $\hat{z}$ direction would obscure this directionality, especially if the averaging distance equals an integer number of periodicities $L_D$ or is very long. However, as we will explore in the following section, surface spirals(SS) emerge within the superficial layers, primarily interacting with the conical spiral near the film surfaces. This interaction accentuates the intrinsic in-plane directionality of the cone.

The in-plane directionality is encoded in the reorientation angle $\theta$ measured right at the surface.  
We always translate the origin so that the surface under examination—be it the top or the bottom—lies at $z=0$; there, $\theta$ equals the phase factor $\phi_0$.  
Throughout the paper we consistently use $\phi_0$ to denote this directionality for each surface separately without further notice.

By letting $A=L_D^2 H_D \ms/(8\pi^2)$ and $|D|=L_D H_D \ms/(2\pi)$, \req{eq:energy} becomes dimensionless if we express the length in unit of $L_D$, external field in units of $H_D$, and energy density in units of $\mu_0 \ms H_D$. We use a cuboid distribution of spin sites with the distance between neighboring sites $d_{x,y}=L_D/16$ along the $\hat{x}$ and $\hat{y}$ directions and a smaller distance $d_z=L_D/64$ along the $\hat{z}$ direction. We also carried out the calculations using a finer mesh ($d_{x,y}=L_D/64$ and $d_z=L_D/128$) to ensure sufficient convergence. Periodic boundary conditions are employed along the $\hat{x}$ and the $\hat{y}$ directions, whereas the $\hat{z}$ direction primarily utilizes free boundary conditions. 

To investigate the magnetic textures arising at the interfaces of different domains, we place two stabilized domains adjacent to each other and then relax the system by numerically integrating the Landau-Lifshitz-Gilbert (LLG) equation:
\begin{equation}
  \frac{\ud \bfm}{\ud \tau} = -\bfm \times \bfh_\text{eff} -\alpha\bfm\times(\bfm\times \bfh_\text{eff}).
  \label{eq:llg}
\end{equation}
Here, $\bfm=\bfM/M_s$, $\tau$ is the effective time, $\alpha$ is the damping constant, and $\bfh_\text{eff}=\bfH_\text{eff}/M_s= (\partial E/ \partial \mathbf{m})/(\mu_0 M_s^2)$ is the dimensionless effective field. To accelerate the minimization process, we disable the non-energy-consuming precession term. The simulation is executed using our self-developed GPU-accelerated code \cite{kim2020mechanisms_,kim2025Topological_,kim2021Kinetics_}. 

\begin{figure}[tpb]
  \includegraphics[width=0.99\linewidth,clip,angle=0]{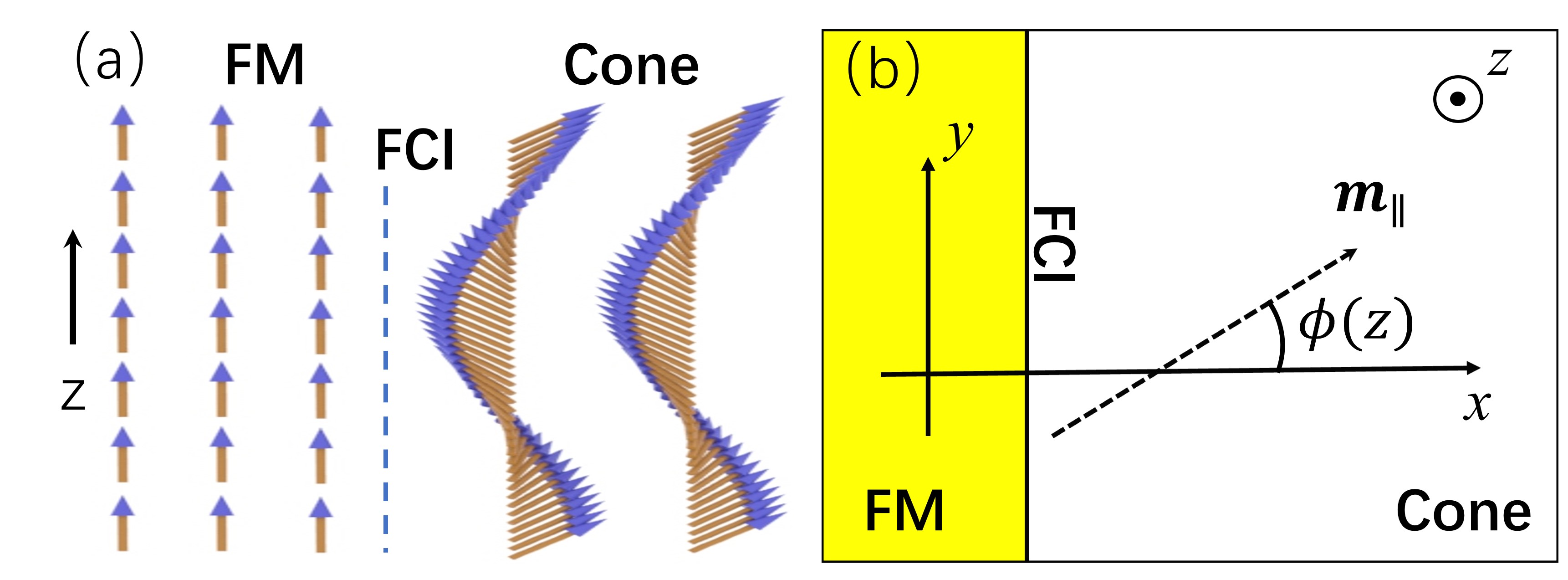}
  \caption{(a) An FCI comprising of a FM domain (left section) and a cone domain (right section). (b) The in-plane directionality of the cone characterized by the angle $\phi(z)$ between the normal vector of the FCI (aligned with the $\hat{x}$ axis), and the in-plane projection of the magnetization vector of the cone, denoted as $\bfm_{{\it \parallel}}=(m_x,m_y,0)=(\cos\phi(z),\sin\phi(z),0)$.}
  \label{fig:0}
\end{figure}

\section{Results}
\subsection{Surface Spirals Formed Near Ferromagnetic-Cone Interface}
In this section, we investigate in-plane conical phase dimensionality controlled magnetic textures that emerge at the surface near FCI. We place a FM domain within a conical domain that is defined by Eq. (\ref{eq:m}), which incorporates a specific external field $H<H_D$ and in-plane directionality $\phi_0$. The magnetization direction in the FM domains is aligned with the applied external field. To stabilize this alignment, a higher field of $H=2H_D$ is applied within its region. Following this, we numerically minimize the system's energy to attain stable states.
\subsubsection{Surface Spirals Formed Near 1D Flat Ferromagnetic-Cone Interface}

Our analysis begins with a 1D flat FCI where the normal direction aligns with the $\hat{x}$-axis. By enforcing spin parallelism along the $\hat{y}$-direction, we effectively reduce the system's dimensionality from 3D to 2D, confining the FCI to a line within the $x$-$z$ plane (Fig.~1(a-b)). 
To avoid interplay between SSs arising at bottom and top surfaces, we consider thick samples with thickness $L_z=8L_D$ much larger than the penetration length of the SS ($\sim0.5L_D$).
Along the $\hat{x}$ direction, the central region is designated as the FM domain, while the remaining regions are modeled as conical domains. As shown in the following, the non-conical structure (i.e., SSs) extends at most to $8L_D$ from the FCI. Both the ferromagnetic domain ($8L_D$ width) and surrounding conical domains ($24L_D$ total width) possess sufficient spatial extent to effectively emulate infinite-width characteristics.

The conical chirality is set by the sign of~$D$.  
Spatial inversion reverses both the chirality and the interface normal~$\mathbf{n}\,(=\pm\hat{x})$, so we simulate only $D<0$ with $\mathbf{n}=+\hat{x}$ (right) and $-\hat{x}$ (left); results for $D>0$ follow from swapping left and right.

\begin{figure}
  \begin{center}
    \includegraphics[width=0.99\linewidth]{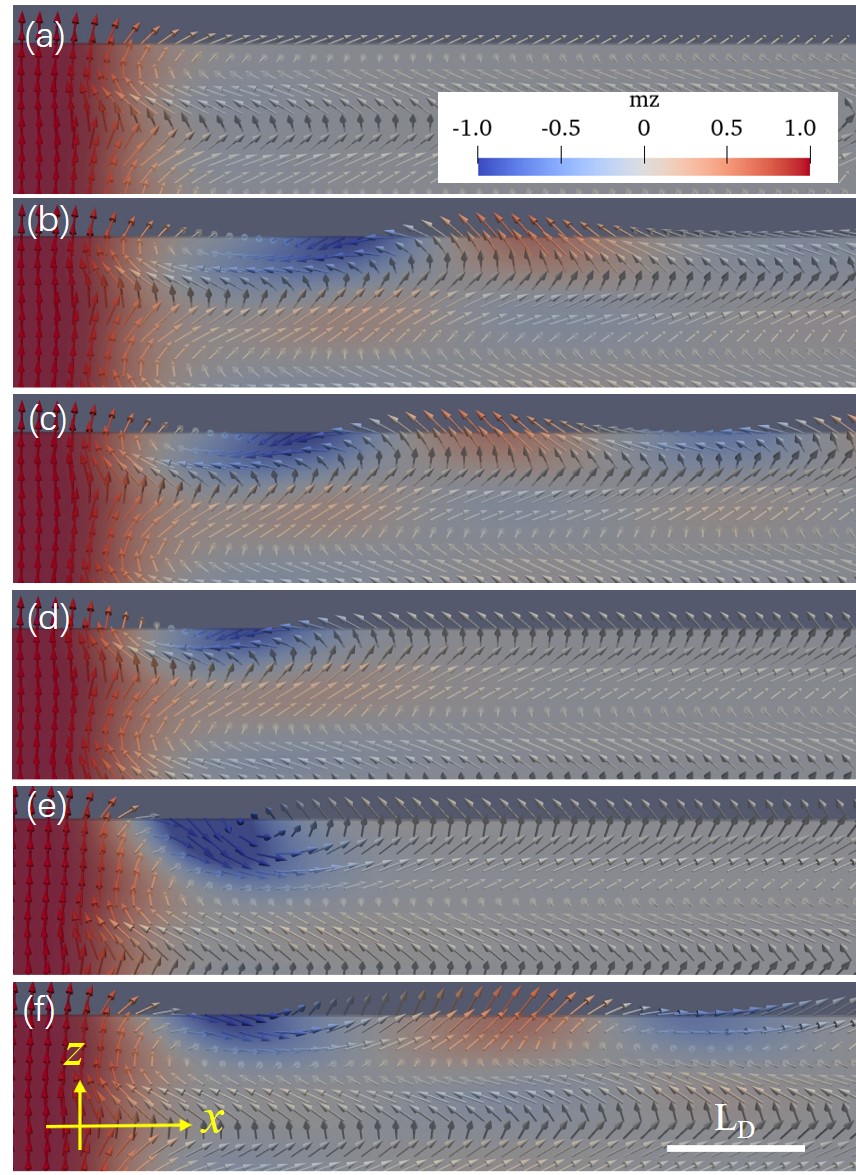}\\
  \end{center}
  \caption{Magnetic structure of SSs spontaneously formed near the FCI, illustrated for the top-right quadrant at a field $H/H_D=0.41$. The structures are shown for varying surface phase factor: (a) $\phi_0=0$, (b) $\phi_0=0.53\pi$, (c) $\phi_0=0.66\pi$, (d) $\phi_0=\pi$, (e) $\phi_0=1.03\pi$, and (f) $\phi_0=1.59\pi$. Each panel displays only the superficial layers of the top-right quadrant to clearly resolve the structural details. The color scale represents the $z$-component of magnetization $m_z$, and the arrows indicate the in-plane spin orientation.}
  \label{fig:SS}
\end{figure}

In Figure \ref{fig:SS}, we present a diverse array of spontaneously formed magnetic structures simulated on the top surface and right side, under a constant external field of $H=0.41H_D$ and varying in-plane cone directionality parameter $\phi_0$ (The evolution of magnetic textures covering all the simulation cell for varying $\phi_0$ are shown in Mov. (1) in supplementary material~\cite{Supplementary}). Notably, SS confined within superficial layers of thickness approximately $L\sim L_D/2$ emerge for a wide range of $\phi_0$ values, with a notable exception at $\phi_0=0$.

The intricate interplay between the shape and existence of these SSs and the surface in-plane cone directionality is captivating. Specifically, when $\phi_0=0$, no SS forms (Fig. 2(a)). Conversely, at $\phi_0=0.53\pi$, a solitary SS emerges adjacent to the FCI (Fig. 2(b)). As $\phi_0$ increases to $0.66\pi$, a transition occurs to multiple SSs, exhibiting a decay in amplitude as they distance themselves from the FCI (Fig. 2(c)). Interestingly, the formation of this ripple-like multi-SS structure is because, at this specific $\phi_0$ values, a FM-like shell is formed on the other side of the first SS, facilitating the formation of the next SS of smaller size. Notably, the periodicity of this multi-SS configuration is approximately $2L_D$, significantly exceeding that of the helical phase.

It is worth emphasizing that the stacked spirals reported in \cite{rybakov2016new_} can be viewed as a manifestation of multi-SSs covering the entire surface of the cone. Their formation can thus be interpreted as the alternating generation of SSs and FM shells. The ability to form multi-SSs further indicates that SSs are not limited to regions near the FCI, but represent a general surface phenomenon that can also occur on the surfaces of a confined conical phase.

At $\phi_0=\pi$, the multi-SSs pattern transforms back into a single SS (Fig. 2(d)). However, upon slightly increasing $\phi_0$ to $1.03\pi$, a distinct single SS emerges, characterized by a deeper penetration length and altered bending directions (Fig. 2(e)).
This change in bending direction
can be interpreted as a reversal of the propagation direction, analogous to the behavior observed in stacked spirals\cite{rybakov2016new_}. Alternatively, it can also be described by a sign change in the topological number, a phenomenon that will be explored in greater detail later.  Finally, at $\phi_0=1.59\pi$, another multi-SSs state arises (Fig. 2(f)), yet its propagation direction is reversed compared to that in Fig. 2(c), highlighting the rich and tunable nature of these magnetic structures.
This rich phase-factor-dependent reconfigurability—enabling four distinct, controllable states within a single device geometry—offers significant potential for high-density polymorphic magnetic storage (HD-PMS) systems. As detailed later, such multi-state switching achieves 2-bit/cell storage without structural modifications, potentially doubling the information density of conventional binary magnetic memory technologies.

\begin{figure}
    \begin{center}
    \includegraphics[width=0.99\linewidth]{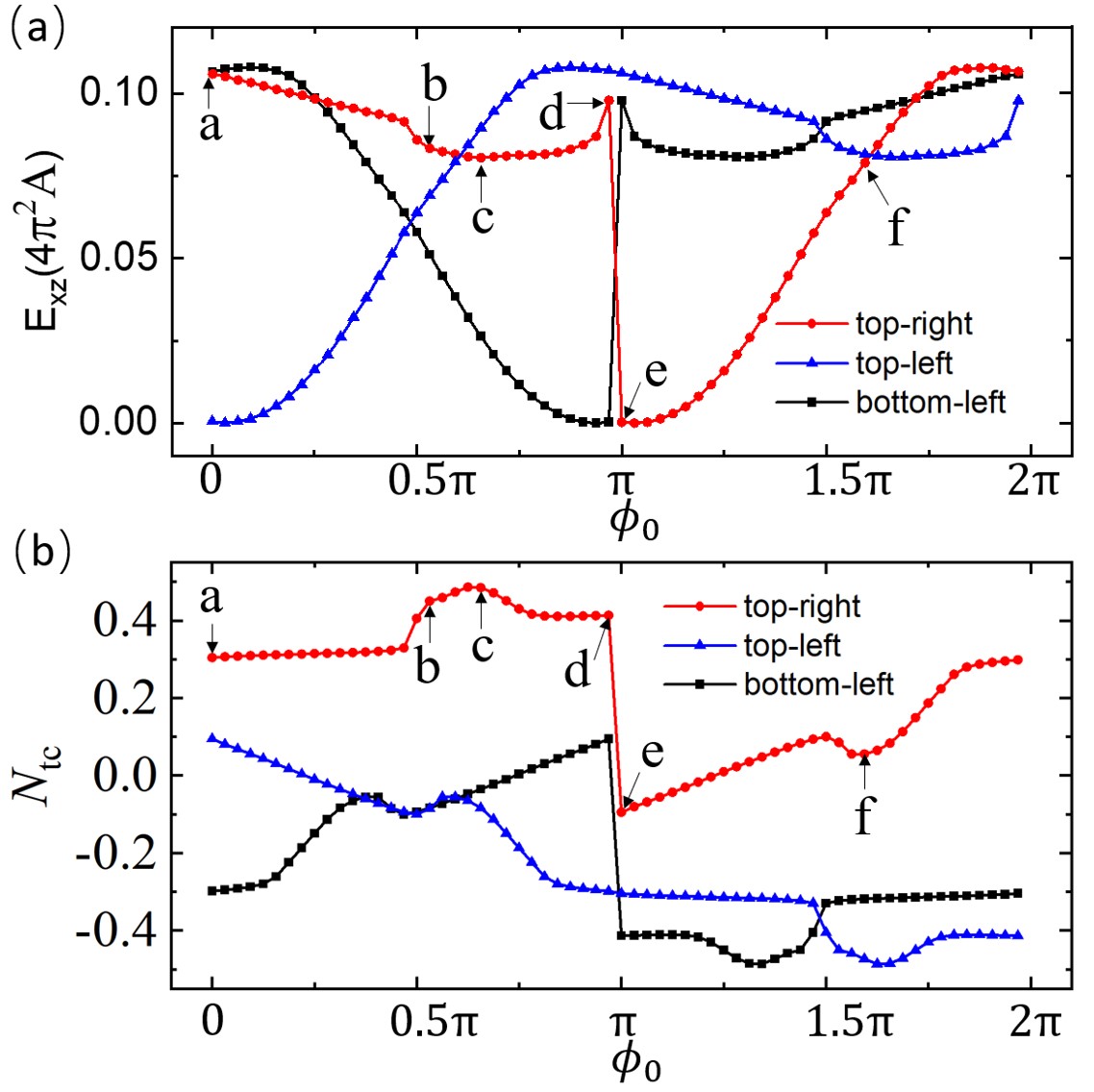}\\
   
    \end{center}
    \caption{ (a) Energy line density \(E_{xz}\) and (b) topological charge \(N_{\text{tc}}\) versus the surface phase factor \(\phi_0\) at \(H/H_D = 0.41\). 
Data are presented for the top-right (red circles), top-left (blue triangles), and bottom-left (black squares) quadrants. 
The curves for the top-left and bottom-left quadrants follow the symmetry relations given by Eqs.~(7)--(10), as described in the text.
Selected points corresponding to the magnetic configurations shown in panels (a) to (f) of Fig.~\ref{fig:SS} are marked with labels a to f.
}
    \label{fig:Topological}
  \end{figure}

  \begin{figure}
    \begin{center}
      \includegraphics[width=0.98\linewidth]{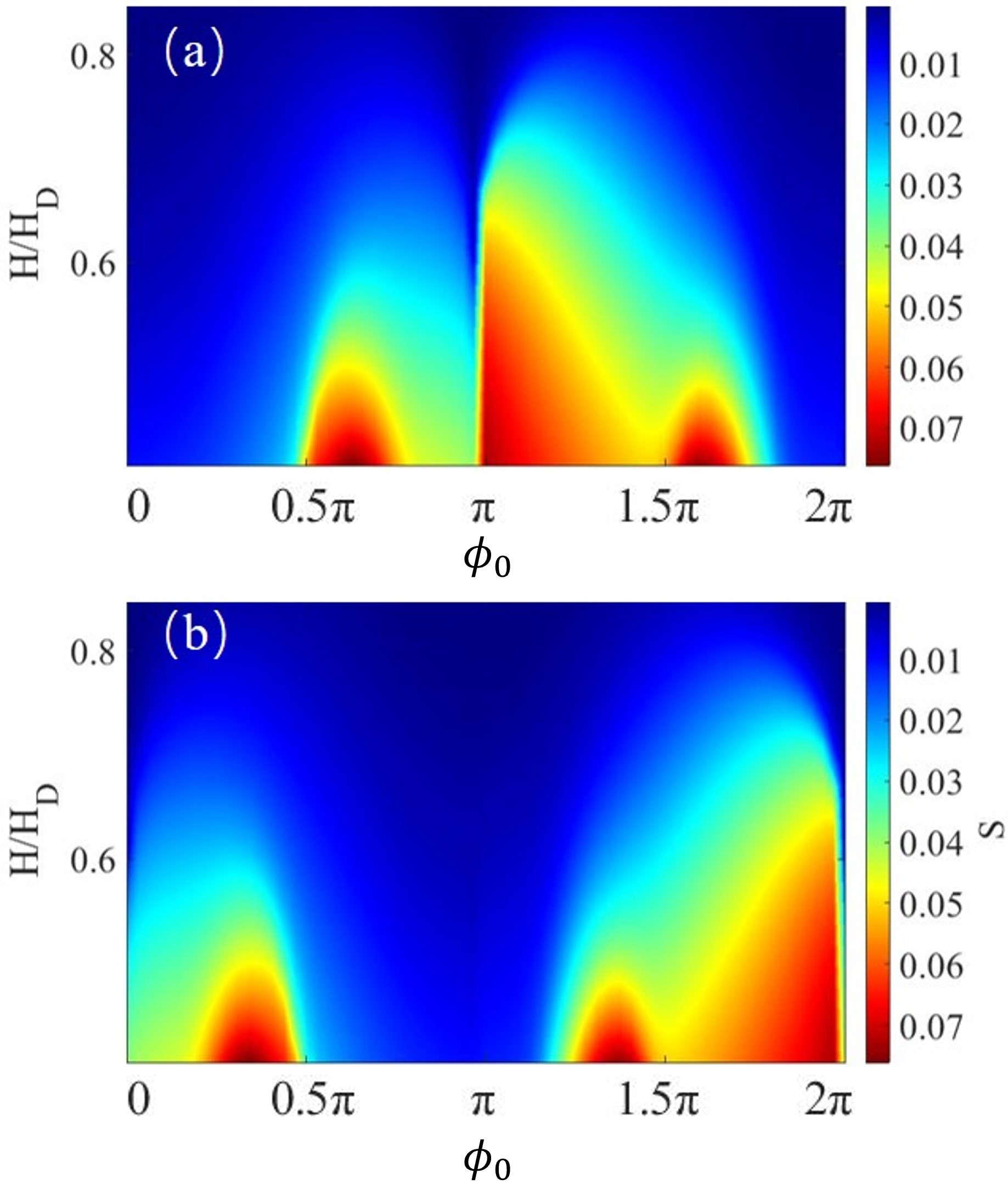}\\
    \end{center}
    \caption{The order parameter $S$, characterizing the size of SS formed at top-right (a) or bottom-right (b) quadrant of the film, as a function of external field $H$ and phase factor $\phi_0$.}
    \label{fig:order_p_S}
  \end{figure}

\begin{figure*}
  \begin{center}
    \includegraphics[width=.98\linewidth,clip,angle=0]{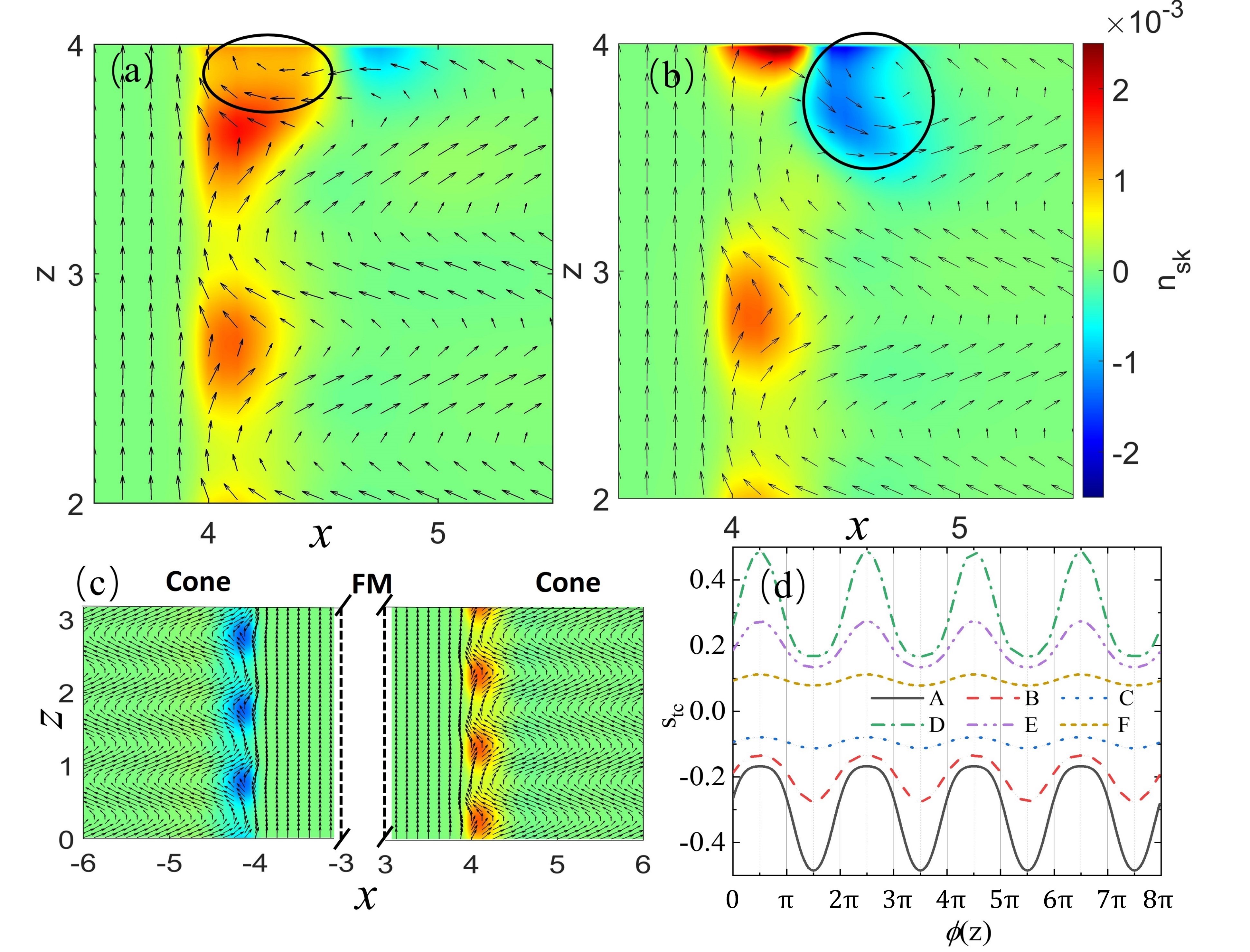} \\
  \end{center}%
  \caption{Topological charge distribution $n_\text{tc}(x,z)$ before (a) and after (b) the discontinuity at $\phi_0\approx\pi$, corresponding to \rfig{fig:SS}(d) and \rfig{fig:SS}(e) respectively. The color plots represent $n_\text{tc}$; arrows show spin orientation projected in x-z plane; black circles mark the positions of SS. (c) Magnetic structure of inner-twists next to 1D FCI in an infinite sample at $H/H_D=0.41$. (d) Topological charge per unit length $s_\text{tc}$ as a function of reorientation angle $\phi(z)$ for the left-side FCI A to C and the right-side FCI D to F. The applied field $H/H_D=0.41$ for A\&D, $0.61$ for B\&E, and $0.81$ for C\&F.}
  \label{fig:ss_size2}
\end{figure*}

\emph{Energy Density and Topological Charge}
The simulation domain is partitioned into four distinct quadrants for detailed analysis: top-right (TR), top-left (TL), bottom-right (BR), and bottom-left (BL). This division is defined by two mutually perpendicular symmetry planes: the horizontal mid-plane of the film (perpendicular to the \(z\)-axis) and the vertical central plane of the FM domain (perpendicular to the \(x\)-axis). 
The symmetry along \(x\) is broken by the DMI, while the symmetry along \(z\) is broken by the external magnetic field. As a result, each quadrant exhibits distinct physical properties. In the following analysis, we quantitatively characterize the SSs—including their energy and topological number—by integrating relevant quantities over each quadrant.

\textbf{Eenergy line-density:} 
To characterize the energy of SSs formed at different phase factors \(\phi_0\), we define the following energy line-density deviation:

\begin{eqnarray}
E_{xz} &=& \iint \left[ E(\phi_0, x, z) - E_{\text{min}} \right]  \mathrm{d} x  \mathrm{d} z,
\label{eq:E_xz}
\end{eqnarray}

where the integration is performed over one of the four quadrants. Here, \(E(\phi_0, x, z)\) is the energy density given by Eq. (1) for a state stabilized at \(\phi_0\), and \(E_{\text{min}}\) is a constant reference energy density chosen such that the minimum of \(E_{xz}\)—corresponding to the lowest-energy state among all \(\phi_0\) configurations—is shifted to zero. Subtracting this baseline renders \(E_{xz}\) independent of the simulation cell size for sufficiently large cells, since expanding the integration region adds equal contributions from \(\phi_0\)-independent uniform phases, which cancel in the difference. Thus, \(E_{xz}\) quantifies the energy increase resulting from changes in the SS structure, as illustrated in Fig.~\ref{fig:SS}, due to variations in \(\phi_0\).

To express the results in dimensionless form, we scale \(E_{xz}\) by \(|E_{\text{helix}}| \cdot L_D^2 = 4\pi^2 A\), where \(E_{\text{helix}} = -D^2/(4A)\) is the bulk energy density of a perfect helical phase at zero field \cite{kim2025Topological_}.

\textbf{Topological charge:} 
The topological charge in the \(x\)-\(z\) plane is defined as:
\begin{eqnarray}
  N_\text{tc} &=& \iint n_\text{tc}(x,z) \, \mathrm{d} x \, \mathrm{d} z \nonumber\\
  &=& \iint \mathbf{m} \cdot \left( \frac{\partial \mathbf{m}}{\partial z} \times \frac{\partial \mathbf{m}}{\partial x} \right)  \mathrm{d} x \, \mathrm{d} z,
  \label{eq:N_tc}
\end{eqnarray}
where \(n_\text{tc}\) denotes the two-dimensional topological charge density, and the integral is evaluated separately over the TR, TL, BR, and BL regions.

\textbf{Eenergy line-density and topological charge  vs. phase factor:} 
Figure 3(a) and (b) shows the evolution of \(E_{xz}\) and \(N_\text{tc}\) as functions of \(\phi_0\) in different quadrants.
In the top-right quadrant, both \(E_{xz}\) and \(N_\text{tc}\) exhibit a complex dependence on \(\phi_0\), consisting of linear segments, a peak, and a discontinuity. Additionally, \(N_\text{tc}\) features a valley. As \(\phi_0\) increases from zero, \(E_{xz}\) decreases linearly at a slow rate, while \(N_\text{tc}\) increases linearly. This behavior corresponds to the region where no SSs are formed, as shown in Fig. 2(a). A sharp drop in \(E_{xz}\) and a rise in \(N_\text{tc}\) occur at \(\phi_0 = 0.469\pi\), corresponding to the emergence of a positively charged SS (Fig. 2(b)). The topological charge reaches a maximum at \(\phi_0 = 0.625\pi\), marking the transition to a multi-SS configuration (Fig. 2(c)). As \(\phi_0\) increases further, the multi-SS state transforms back into a single SS (Fig. 2(d)), causing \(N_\text{tc}^{\text{TR}}\) to decrease. In this region, \(E_{xz}\) remains nearly constant. A sharp drop in both \(E_{xz}\) and \(N_\text{tc}\) occurs near \(\phi_0 = \pi\). The decrease in \(N_\text{tc}\) signifies a reversal in the bending direction of the SS (compare Fig. 2(d) and (e)), which also inverts its topological charge and thus reduces \(N_\text{tc}\). For \(E_{xz}\), the drop is attributed to the larger size of the SS after the transition (Fig. 2(e)) compared to that before the transition (Fig. 2(d)). Note that a small increase in \(E_{xz}\) precedes the sharp drop.

As \(\phi_0\) continues to increase, the negatively charged SS gradually shrinks, leading to a rise in both \(E_{xz}\) and \(N_\text{tc}\). The \(N_\text{tc}\) curve exhibits an additional valley at \(\phi_0 = 1.59\pi\), indicating the emergence of a negatively charged multi-SS state (Fig. 2(f)). When \(\phi_0\) approaches \(2\pi\), which is equivalent to \(\phi_0 = 0\), both \(E_{xz}\) and \(N_\text{tc}\) return to their initial values, completing a full cycle of topological charge variation. Notably, \(N_\text{tc}\) remains predominantly positive due to contributions from positively charged inner-twists, which will be discussed in detail later.

\textbf{Energy per bit estimation:} The energy differences between difference SS states, as observed in Fig. 3(a), are $<0.1\times 4\pi^2 A \times L_y$. For typical material parameters (exchange stiffness $A \approx 1\times10^{-11}$ J/m) and a device width of 500 nm along the $y$-direction, this corresponds to a switching energy  on the order of  $10^{-17}$ J per bit. This low energy requirement highlights the potential for energy-efficient SS-based memory devices.

\textbf{Symmetry relations:} 
We denote the energy and topological charge integrated over a specific quadrant with corresponding superscripts, such as $E_{xz}^{\text{tr}}$ and $N_{\text{tc}}^{\text{tr}}$ for the top-right region. The configurations in other quadrants can be derived from the top-right quadrant through rotations and reflections. For instance, the left quadrant is obtained by rotating the right quadrant by 180$^\circ$ about the $z$-axis, which corresponds to a phase shift of $\pi$ in the reorientation angle. This yields the relation for the energy:
\begin{equation}\label{eq:E_tl}
  E_{xz}^{\text{tl}}(\phi_0) = E_{xz}^{\text{tr}}(\phi_0 + \pi).
\end{equation}
For the topological charge $N_\text{tc}$, the rotation also  inverses the $\hat{z} \times \hat{x}$ direction (i.e., the $\hat{y}$-axis),  therefore, the sign of $N_\text{tc}$ is reversed. Consequently, the topological charges in the left and right quadrants are related by:
\begin{equation}\label{eq:N_tl}
  N_{\text{tc}}^{\text{tl}}(\phi_0) = -N_{\text{tc}}^{\text{tr}}(\phi_0 + \pi).
\end{equation}

Similarly, the bottom quadrants are related to the top ones by reflection. For the bottom-left quadrant, we have:
\begin{equation}\label{eq:E_bl}
  E_{xz}^{\text{bl}}(\phi_0) = E_{xz}^{\text{tr}}(-\phi_0),
\end{equation}
\begin{equation}\label{eq:N_bl}
  N_{\text{tc}}^{\text{bl}}(\phi_0) = -N_{\text{tc}}^{\text{tr}}(-\phi_0).
\end{equation}
Finally, the bottom-right quadrant can be expressed in terms of the top-left and top-right quadrants as:
\begin{equation}\label{eq:E_br}
  E_{xz}^{\text{br}}(\phi_0) = E_{xz}^{\text{tl}}(-\phi_0) = E_{xz}^{\text{tr}}(\pi - \phi_0).
\end{equation}
\begin{equation}\label{eq:N_br}
  N_{\text{tc}}^{\text{br}}(\phi_0) = -N_{\text{tc}}^{\text{tl}}(-\phi_0) = N_\text{tc}^{\text{tr}}(\pi - \phi_0).
\end{equation}

In Figs. 3(a) and 3(b), we present \(E_{xz}\) and \(N_{\mathrm{tc}}\) for the top-right, top-left, and bottom-left quadrants. The data consistently satisfy the symmetry relations given by Eqs.~\eqref{eq:E_tl}--\eqref{eq:N_bl}. For visual clarity, the bottom-right quadrant is omitted from these plots, as its behavior can be directly derived from the presented results using Eqs.~\eqref{eq:E_br} and \eqref{eq:N_br}.

\emph{Quantitative Evaluation} The size of SSs can be measured using the topological charge discussed above. However, the cancellation of positive and negative charges necessitates a more robust order parameter to quantitatively evaluate the dimensions of SSs. To address this, we employ a positive-definite quantity \( S \), where a larger value indicates a more extensive SS structure. This quantity is defined as:
\begin{equation}
  S=\int_{0}^{l_\text{c}^x}\int_{-l_\text{c}^z}^0 -m_z^* \, \Theta(-m_z^* ) \, \ud x\, \ud z,
\end{equation}
where, $\Theta$ is the Heaviside step function, and $m_z^*=m_z-H/H_D$ captures the deviation of $m_z$ from the characteristic value of a pure conical phase. The integration limits $l_\text{c}^x=12L_D$ and $l_\text{c}^z=l_z/2$, along the $\hat{x}$ and $\hat{z}$ directions, respectively, are set to encompass the relevant region of interest.

\rFig{fig:order_p_S}(a) presents a contour plot illustrating the variation of $S$ as a function of $\phi_0$ and $H$ for the top-right quadrant of the parameter space. For the remaining quadrants, analogous plots can be derived using the transformations outlined in Eqs. (10) to (12). \rFig{fig:order_p_S}(b) gives an example for the bottom-right quadrant. In general, an increase in $H$ leads to a decrease in $S$, whereas its dependence on $\phi_0$ exhibits a non-monotonic behavior. Notably, when $\phi_0\approx0$, $S\approx0$, which is consistent with the observations in \rfig{fig:SS}(a). Two distinct dome-shaped regions with pronounced $S$ values emerge at $\phi_0\approx 0.6\pi$ and $\phi_0\approx 1.6\pi$, corresponding to positively and negatively charged multi-SS states (\rfig{fig:SS}(d) and \rfig{fig:SS}(f), respectively). These states, despite not achieving the deepest penetration depth, exhibit larger sizes due to the increased number of SSs. Analogous to the topological charge, $S$ also undergoes a discontinuity at $\phi_0 \approx \pi$.

\emph{Underlying Mechanism} To gain insight into the underlying mechanism responsible for the dramatic change in the SS structure and the resulting discontinuity, we examine the topological number density distribution $n_\text{tc}(x,z)$ near the discontinuity point, as shown in \rfig{fig:ss_size2}(a)--(b). We find that both SSs around the discontinuity interact with a positively charged twist adjacent to the FCI. One spiral merges with the twist, sharing the same charge, while the other, adjacent to it, carries an opposite charge. This interplay accounts for the discontinuity observed in $N_\text{tc}$ and $S$.
Further investigation reveals that these positively charged twists are not confined to the superficial layers but also exist in the inner regions without surface twists, prompting us to classify them as inner-twists. In Fig.~\ref{fig:ss_size2}(c), we illustrate the static configuration of an infinitely thick sample devoid of surfaces, showcasing periodic twists within each cone period. The inner-twists on either side possess opposite charges and are positioned with a half-period shift.

By integrating \( N_\text{tc}(x,z) \) along the \(\hat{x}\) direction, we derive the topological charge line distribution \( s_\text{tc}(z) \) along the \(\hat{z}\) direction. In Fig.~\ref{fig:ss_size2}(d), \( s_\text{tc}(z) \) as a function of \( \phi(z) \) for various \( H \) values resembles sine functions with broadened valleys. The peaks on both sides occur when \( \mathbf{n} \cdot \mathbf{m} \) is maximized (where \( \mathbf{n} \) denotes the FCI's normal direction and \( \mathbf{m} \) the cone's magnetization reorientation). These twists gradually diminish as the cone's magnetization reorientation fades, occurring as the field approaches the saturation field.

Returning to the discontinuity of the SS discussed above, for the phase factor \( \phi_0 = 0 \), \( \mathbf{n} \cdot \mathbf{m} \) is maximized, meaning this layer would be the center of an inner-twist if it were not the surface layer. This implies that the surface layer is exactly at the balance point, where the surface attempts either to broaden (as shown in Fig.~\ref{fig:SS}(a)) or narrow (as shown in Fig.~\ref{fig:SS}(b)) the inner twist. Ultimately, this surface-induced modulation results in two distinct SS structures, which in turn give rise to the observed discontinuity. 

\begin{figure}
  \begin{center}
    \includegraphics[width=0.98\linewidth]{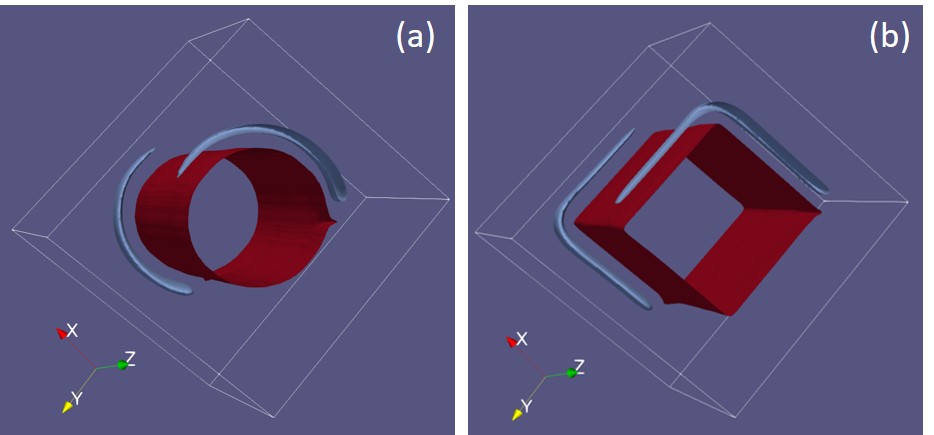}\\
  \end{center}
  \caption{SSs formed near a two-dimensional FCI at $H/H_D=0.41$, created by embedding a ferromagnetic (a) cylinder and (b) square prism into a conical phase. The white and red surfaces represent isosurfaces of the magnetization $z$-component corresponding to $m_z = 0$ and $m_z = 0.99$, respectively.}
  \label{fig:ss_shape}
\end{figure}
\subsubsection{Surface Spirals Formed at 2D Ferromagnetic-Cone Interface}
In this section, we study the SS that emerge in proximity to a two-dimensional FCI. Specifically, we consider two archetypal geometries of the FM domain embedded within a conical phase: a FM square prism and a FM cylinder.

For the cylindrical FCI, an asymmetric eyebrow-shaped SS emerges, characterized by a broader and deeper profile on one side, gradually tapering off on the other (see \rfig{fig:ss_shape} (a)). This observation aligns with our findings for 1D FCIs, where each small segment of the cylinder can be approximated as a 1D FCI. The rotation of the cylinder's normal direction, $\mathbf{n}$ is equivalent to change the cone directionality. Consequently, the varying size of the SS as $\mathbf{n}$ rotates gives rise to the eyebrow-like shape. When the cone is screwed by varying $\phi_0$, the eyebrow rotates due to the change in the cone's directionality. Therefore, changing the film thickness can be seen as fixing the directionality of one surface and screwing the other, which in turn varies the angle between the reorientation of the eyebrow at the top and bottom. Therefore, this angle is exclusively controlled by the film thickness, analogous to the behavior observed in stacked spirals\cite{rybakov2015New_}.

When it comes to the square-prism-shaped FCI, its four facets possess distinct normal directions. Treating each facet as an independent one-dimensional FCI, switching between neighboring facets amounts to a phase shift of $\pi/2$, akin to selecting spontaneously from four equally distributed angles. Importantly, at least one of these angles will necessarily fall within a region where the formation of SSs is not permissible. As a result, the formation of SSs is observed on two or three facets of the square prism, contingent upon the value of $\phi_0$, the directionality of cone. In \rFig{fig:ss_shape}(b), we show the SSs formation at $\phi_0=0$ and $H/H_D=0.41$. 

The programmable formation of eyebrow-like or facet-specific SS patterns around 2D FM structures demonstrates the feasibility of geometrically controlled magnetic textures for potential use in reconfigurable magnetic circuits. Furthermore, each of these SSs, being finite in length, possesses a finite topological number within the $x-y$ plane. They thus serve as embryonic precursors to magnetic bobbers \cite{rybakov2015New_,zheng2018Experimental_,ran2021Creation_,redies2019Distinct_,kim2020mechanisms_}, which can potentially evolve into fully fledged skyrmions. Our simulations have confirmed this transition, and we anticipate delving deeper into this fascinating topic in our future work.

\begin{figure*}
  \begin{center}
    \includegraphics[width=0.98\linewidth]{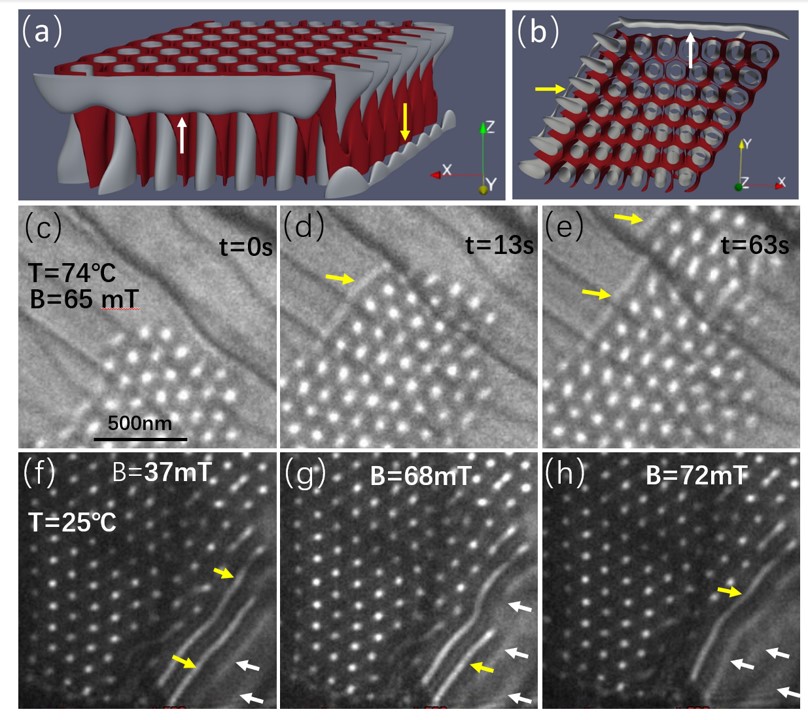}\\
    \caption{ Top (a) and side (b) views of the simulated magnetic structure of a skyrmion cluster formed in a conical phase at \( H = 0.41 H_D \) and thickness \( L_Z = L_D \). The red and white surfaces represent isosurfaces of the magnetization \(z\)-component at \( m_z = 0.95 \) and \( m_z = -0.05 \), respectively. White and yellow arrows indicate SSs formed adjacent to the skyrmion cluster.
(c-h) Experimental LTEM observations of skyrmion clusters in Co\(_8\)Zn\(_{10}\)Mn\(_2\) thin films with thickness \( L_Z = 150 \, \text{nm} \approx L_D \). (c-e) Sequence of images showing the thermal activation and growth of skyrmions in the conical phase, acquired at \( 1.56 \, \text{mm} \) under-focused conditions. These images were obtained at a magnetic field of \SI{65}{mT} ($\approx 0.32 H_D$) and temperature of \SI{74}{\degreeCelsius}. (f-h) Sequence of images captured during rapid field increase at lower temperature, acquired at \( 1.49 \, \text{mm} \) under-focused conditions. The applied magnetic fields are (f) \SI{37}{mT} ($\approx 0.18 H_D$), (g) \SI{68}{mT} ($\approx 0.34 H_D$), and (h) \SI{72}{mT} ($\approx 0.36 H_D$) at a temperature of \SI{25}{\degreeCelsius}.
In panels (c-h), yellow arrows mark high-contrast stripes that may correspond to either helices or SSs, while white arrows indicate low-contrast stripes that are more likely to be SSs due to their characteristic appearance.}
    \label{fig:exp}
  \end{center}
\end{figure*}

\subsection{Surface Spirals Formed at Edges of a Skyrmion Cluster Merged in Conical Phase}
Experimentally, creating artificial FCIs can present significant challenges. However, the intricate twisting configurations inherent in chiral magnets offer a promising avenue for generating localized FM domains with subtle effects, thus providing an alternative platform for proving and applying the finding we presented in the above section. The above-mentioned multi-stacked-spiral state is already an example. The formal SS generates an FM-like shell next to it, which promotes formation of the next SS. For skyrmions merged in conical phase, they are enveloped by FM shells. Notably, an isolated skyrmion has a tube-like core and a spring-like FM shell winding around the core\cite{kim2020mechanisms_}. The partially covered FM shell gradually becomes fully covered when it coupled to other skyrmions. For a skyrmion fully surrounded by six neighboring skyrmions, i.e., inner skyrmions within a skyrmion cluster, it is fully covered by a FM shell. For the skyrmions at the edges of the cluster, they are partially covered, depending on the directionality of cone refer to the normal direction of the cluster facet. To promote formation of FM shell and SSs, a suitable directionality is required that can support both the formation of FM shells and SS. The FM shells are formed when $\phi_0\approx \pi$, which, as shown in Fig. \ref{fig:Topological} and \ref{fig:order_p_S}, also supports formation of SSs.

To confirm the above hypothesis, we initiated our investigation from a conical phase and artificially introduced several skyrmions, arranging them to form skyrmion clusters. The stable configuration was determined by minimizing the system energy. Figures~\ref{fig:exp}(a) and (b) present the numerically calculated magnetic configuration of a rhombus-shaped skyrmion cluster formed in a conical phase at \( H = 0.41H_D \). FM shells are formed, partially covering the facets of the rhombus. A SS is observed adjacent to the FM envelope in the vicinity of each surface. The reorientation angle is \(\frac{3}{2}\pi\) for the top surface and \(\frac{7}{6}\pi\) for the bottom surface. According to Fig.~\ref{fig:order_p_S}(a), the former results in a large SS, whereas, as shown in Fig.~\ref{fig:order_p_S}(b), the latter leads to a small SS. This expectation is entirely aligned with the result shown in Figs.~\ref{fig:exp}(a) and (b), where the SS in the vicinity of the top surface is much smaller than that in the vicinity of the bottom surface.

We further conducted LTEM observations of skyrmion clusters in Co$_8$Zn$_{10}$Mn$_2$ thin films with a thickness of approximately \SI{150}{nm} (\(\sim L_D\)). To prepare the sample, a bulk Co$_8$Zn$_{10}$Mn$_2$ sample, with a critical temperature \( T_c \sim \SI{370}{K} \), was synthesized by sealing individual high-purity metals (all \( > 99.9\% \) purity) in a quartz ampule backfilled with ultrahigh-purity argon. The ampule was heated to \SI{1000}{\degreeCelsius} for 12 hours, cooled at a rate of \SI{1}{\degreeCelsius/h} to \SI{925}{\degreeCelsius}, held at this temperature for 96 hours, and then quenched in water.

The sample was obtained from a large single crystalline grain (\(\sim \SI{100}{\micro\meter}\)) free of grain boundaries. It was mounted on a MEMS heater using a focused ion beam. LTEM observations were performed using an FEI Titan Themis microscope equipped with an FEI NanoEx-i/v in-situ TEM holder, enabling precise temperature control under isothermal conditions. An external magnetic field was applied along the electron beam direction by partially exciting the objective lens. LTEM videos were recorded at 20 frames per second using an FEI Ceta camera to capture the evolution of magnetic textures (see Ref.~\onlinecite{kim2020mechanisms_} for further details on the experimental setup).

By manipulating temperature and external magnetic fields, we observe transitions between different magnetic states, including helical phase, skyrmion lattices, conical phase, and FM phase (see the phase diagram in Fig.~S1(d) of Ref.~\cite{kim2020mechanisms_}).

As shown in our simulations (Fig.~\ref{fig:exp}(a-b)), SSs are half-penetrated metastable objects, structurally similar to magnetic bobbers \cite{rybakov2015New_,zheng2018Experimental_,ran2021Creation_,redies2019Distinct_,kim2020mechanisms_}. With thermal fluctuation assistance, SSs may either fully penetrate the film and transform into helices or become dissipated. To experimentally capture these transient states, one must examine partially relaxed transition states rather than fully relaxed ground states composed of uniform helical, skyrmion, or conical phases.

In Fig.~\ref{fig:exp}(c-e), we present a sequence of images captured at a fixed field of \SI{65}{mT} ($\approx 0.32 H_D$) and elevated temperature of \SI{74}{\degreeCelsius} near the Curie temperature (\SI{97}{\degreeCelsius}), showing the temporal evolution of skyrmion growth from the conical phase. Figure~\ref{fig:exp}(c) shows a partial skyrmion cluster that grows over time, with a single stripe emerging on the newly grown facet of the cluster (Fig.~\ref{fig:exp}(d)). Continued growth causes the stripe to elongate with the expanding cluster until the cluster develops a step-like feature, forming two parallel facets. The stripe subsequently breaks and continues to grow on the newly formed parallel facets.

Given the thin experimental sample and relatively high temperature, SSs can readily penetrate the film fully via thermal activation. The observed stripes in Fig.~\ref{fig:exp}(d-e) could represent either SSs or helices evolved from SSs, with their lower contrast compared to skyrmions suggesting a higher probability of SS identity. These stripes decorate either single facets (Fig.~\ref{fig:exp}(d)) or parallel facet pairs (Fig.~\ref{fig:exp}(c)), consistent with our simulations (Fig.~\ref{fig:exp}(a-b)) and the theoretical understanding that SS stability is maximized at an optimal angle between facet normal and in-plane cone orientation. Since the in-plane cone orientation is fixed experimentally (or slowly varies due to thermal fluctuation), SSs selectively appear on facets with the most favorable geometric configuration.

In Fig.~\ref{fig:exp}(f-h), we show image sequences acquired at a significantly lower temperature of \SI{25}{\degreeCelsius} under rapidly increasing magnetic fields (starting from zero). Lower temperatures stabilize more metastable states. As the field increases from zero, helices gradually transform into conical phase or skyrmions. Figure~\ref{fig:exp}(f) shows two high-contrast surviving helices decorating a skyrmion cluster: one immediately adjacent to the cluster and another next to the first. Beyond these, faint low-contrast stripes with larger periodicity than the inner two are observed, resembling the multi-SS configurations seen in our simulations (Fig. 2 (c) and (f)). When the field increases to \SI{68}{mT} ($\approx 0.34 H_D$), the stripe second-closest to the skyrmion cluster begins to lose contrast from one end, forming a hybrid structure composed of half-SS and half-helix, indicating its transition from helix to SS. Further increasing the field to \SI{72}{mT} ($\approx 0.36 H_D$) completes the transformation of the entire helix into an SS.

Notably, pronounced multi-SS configurations are observed in Fig. ~\ref{fig:exp}(f-h), in contrast to their absence in Fig.~\ref{fig:exp}(c-e). This striking difference originates from the distinct thermodynamic and interfacial environments in the two cases. As indicated in Figs. 2(c) and 2(f), the size of an SS generally decays with increasing distance from the FCI. Smaller SSs, being less stable, are more susceptible to thermal evaporation. In the case of Fig.~\ref{fig:exp}(c-e), the elevated temperature promotes this evaporation. Conversely, in Fig.~\ref{fig:exp}(f-h), the SSs are stabilized by two key factors: the lower temperature itself, and the fact that they are anchored adjacent to a helix rather than a skyrmion cluster. This geometry enables a stronger stripe-to-stripe coupling compared to the weaker stripe-to-skyrmion-cluster interaction. The combined effect of reduced thermal fluctuations and enhanced inter-stripe coupling thus enables the clear observation of multi-SS states in the latter regime.

Apart from the similarity in magnetic texture, several other experimental observations align with our simulations. For instance, our experiments show that there is an upper limit to the magnetic field for observing skyrmion clusters adorned with stripes. This is consistent with our simulations, which show that higher fields lead to the decay of SSs.  

We note that the theoretically predicted optimal field within model Eq.~\eqref{eq:energy} is $H \approx 0.5H_D$ \cite{rybakov2016new_, leonov2016chiral_}. The observed deviation likely originates from neglected contributions such as demagnetization-induced magnetic anisotropy or single-ion anisotropy associated with Mn atoms. These precisely controlled conditions enable reliable capture of metastable SS-adorned skyrmion clusters.

In summary, our combined experimental and simulation results demonstrate that skyrmion clusters in conical phase develop FM shells that facilitate SS formation at their edges. The consistency between LTEM observations and theoretical predictions validates the proposed mechanism of SS stabilization through optimal geometric configurations.

\begin{figure*}
  \centering
  \includegraphics[width=0.98\linewidth]{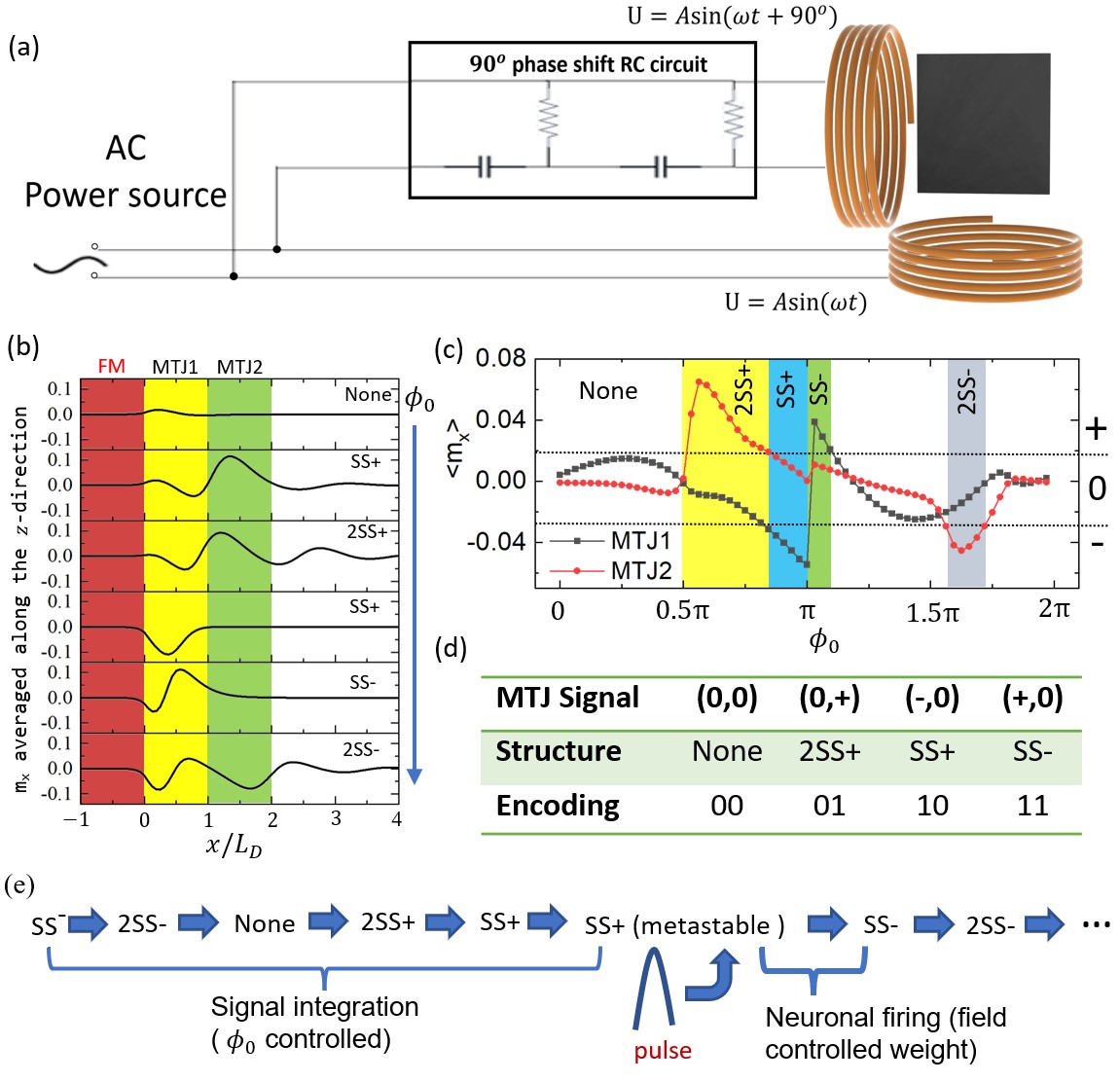}
  \caption{Spintronic applications enabled by phase factor controlled SSs. 
  (a) Schematic diagram of the device concept for generating a circularly polarized AC field to control phase factor $\phi_0$. The practical implementation at GHz frequencies may utilize a microwave resonant cavity. 
  (b) The $m_x$ component, averaged along the $z$-direction, as a function of $x$ in the top-right quadrant. The six curves (top to bottom) correspond to the magnetic configurations displayed in Figs. 2(a-f), respectively: near-zero (no SS), positive/negative peaks (single SS-/SS+), and decaying-wave profiles (multi-SS, marked by 2SS+ and 2SS-). 
The red-shaded region denotes the FM domain, while two MTJ detectors are positioned adjacently in sequence. Their respective detection zones are highlighted by yellow and green shading.
  (c) Average $m_x$ signals from two MTJ detectors versus $\phi_0$, with classification thresholds at $\langle m_x\rangle^{\text{upper}}=0.02$ and $\langle m_x\rangle^{\text{lower}}=-0.03$ (horizontal dotted lines).
   Signals $A_{MTJ1}$/$A_{MTJ2}$ (output of two detectors) are labeled as + (if$> \langle m_x\rangle^{\text{upper}}$), -(if$< \langle m_x\rangle^{\text{lower}}$), or 0 (otherwise).
   Results are ordered as ($A_{MTJ1}$, $A_{MTJ2}$ ).
   Five regions are color-coded: yellow (2SS+, (0,+)), blue (SS+, (-,0)), green (SS-, (+,0)), grey (2SS-, (0,-)), and unshaded (none, (0,0)).
  (d) Binary encoding scheme for the four selected structural states from (c), enabling 2-bit-per-cell storage in HD-PMS.
  (e) Neuromorphic emulation using SS dynamics: Continuous $\phi_0$-driven transitions between SS- and SS+ states simulate neuronal integration, while stimulus-triggered jumps between metastable states replicate firing. Threshold tunability via $\phi_0$ and external field $H$ enables synaptic-like programmability.}
  \label{fig:apply}
\end{figure*}

\section{Potential Spintronic Applications}
There exist several viable approaches to manipulate the phase factor $\phi_0$, including the use of circularly polarized AC fields \cite{delser2021Archimedean_} or magnon currents \cite{wang2019Magnetization_}. As demonstrated in Ref.~\cite{delser2021Archimedean_}, a finite screw velocity $\Omega_{\text{screw}}$ can be induced by an arbitrarily weak driving field in the absence of pinning. The magnitude of $\Omega_{\text{screw}}$ depends on both the amplitude—exhibiting a quadratic dependence on the driving field $B$ at small amplitudes—and the frequency of the driving field, showing a sharp resonant enhancement when the driving frequency matches either of the two characteristic resonance frequencies of the conical phase (which correspond to its left- and right-handed modes and typically lie in the GHz range). Under such resonant driving, the screw motion of the magnetic texture occurs at a characteristic frequency $\Omega_{\text{screw}}/2\pi$ typically in the 1–10 MHz range.

Figure~\ref{fig:apply}(a) schematically illustrates the operating principle for generating a circularly polarized AC field using a pair of orthogonal solenoids driven by currents with a $90^\circ$ phase difference. To efficiently excite the conical phase near its intrinsic GHz-range resonances, a practical implementation of this concept employs a microwave resonant cavity. Such cavities constitute a well-established technique in ferromagnetic resonance experiments for generating strong, high-frequency rotating magnetic fields.

Consequently, the phase factor evolves as
\begin{equation}
\phi_0(t) = \phi_0(0) + \Omega_{\text{screw}} \cdot t,
\end{equation}
enabling precise temporal control via the duration of the applied AC field.

The precise phase control, combined with the rich texture variety of SSs, opens up unprecedented opportunities for spintronic devices. A primary implementation is the SS-based high-density programmable magnetic storage (HD-PMS). Figure~\ref{fig:apply}(b) shows the $x$-component of SS magnetization averaged along the $z$-direction through a depth $L_D$ from the surface in the top-right quadrant, with the six curves corresponding to the magnetic configurations displayed in Figs. 2(a-f). The curves exhibit dramatic differences across different structures: $m_x$ remains nearly zero for states without SSs; a positive/negative peak corresponds to a negatively/positively charged single SS; and decaying-wave-like profiles indicate multi-SS configurations. These distinct features enable the use of two MTJ detectors (MTJ1 and MTJ2), each with width $L_D$, placed adjacent to the FM domain. Figure~\ref{fig:apply}(c) plots the average $m_x$ within the regions of MTJ1 and MTJ2 as functions of $\phi_0$. By setting an upper threshold $\langle m_x\rangle^{\text{upper}}=0.02$ and a lower threshold $\langle m_x\rangle^{\text{lower}}=-0.03$, the detected $\langle m_x\rangle$ values are classified into three categories: those larger than $\langle m_x\rangle^{\text{upper}}$ are marked ``+'', those smaller than $\langle m_x\rangle^{\text{lower}}$ are marked ``-'', and intermediate values are marked ``0''. Combining the classifications from MTJ1 and MTJ2 allows recognition of at least four distinct states (color-coded in Fig.~\ref{fig:apply}(c)), as summarized in Fig.~\ref{fig:apply}(d). Encoding data with these four configurations enables an HD-PMS achieving 2-bit-per-cell storage density. Furthermore, by incorporating finer threshold discrimination (e.g., by adjusting the values of $\langle m_x\rangle^{\text{upper}}$ and $\langle m_x\rangle^{\text{lower}}$), analyzing temporal signal evolution sequences, or integrating signals from bottom SSs, the system could potentially recognize additional distinct states. This approach paves the way for higher-density multi-bit-per-cell storage architectures, scaling from the current 2-bit (4 states) implementation to 3-bit (8 states) and beyond within the same device footprint.

Regarding its sensitivity to variations, the deliberately set thresholds ($\langle m_x\rangle^{\text{upper}}=0.02$ and $\langle m_x\rangle^{\text{lower}}=-0.03$) provide a wide margin for reliable state discrimination. This design makes the readout scheme robust against signal fluctuations that may arise from variations in sample thickness or the presence of interfacial disorder. The system can reliably function as long as the thickness significantly exceeds the  period of the cone $L_D$, and the disorder scale remains smaller than the spatial extent of an SS (approximately $0.5L_D$ in depth and $L_D$ in width), which ensures the preservation of the distinctive $m_x$ profiles essential for state discrimination.

Building on HD-PMS, the same hybrid SS dynamics facilitate neuromorphic computing \cite{burr2017Neuromorphic_}. As shown in Fig.~\ref{fig:apply}(e), continuous $\phi_0$-driven transitions between SS$^-$ and SS$^+$ states emulate the \textbf{integration phase} of biological neurons, mirroring the gradual accumulation of membrane potential. The abrupt topological transition from metastable SS$^+$ to SS$^-$, triggered at a critical stimulus threshold, replicates the \textbf{firing phase} of action potential generation. Notably, this threshold can be dynamically tuned via $\phi_0$ and external field $H$, offering synaptic-like programmability for adaptive computing.

\textbf{Key Advantage}---Unlike conventional skyrmion-based neurons \cite{chen2018Compact_,li2021Magnetic_a}, which are largely limited to binary ``present/absent'' encoding, SS neurons exhibit an intrinsic continuous-to-discrete duality. Smooth modulation of $\phi_0$ continuously sculpts the SS topological charge, enabling \emph{analog accumulation}, while the ensuing discontinuous jump provides an \emph{all-or-none} spike. Consequently, a single SS element natively supports multi-bit or fully analog neuromorphic states, substantially enriching the computational alphabet without increasing device footprint.

Furthermore, the phase factor $\phi_0$ establishes an additional control dimension complementary to traditional methods, unlocking broader spintronic functionalities including dynamic manipulation of magnon phase and amplitude through $\phi_0$ modulation. These emerging applications demonstrate $\phi_0$'s versatility as a fundamental control parameter for advanced spintronic systems.

\section{Conclusions}

In conclusion, we have systematically investigated the in-plane directionality of conical phases in surface-confined magnetic structures. Our findings demonstrate the emergence of surface spirals (SSs) at ferromagnetic-conical interfaces (FCIs) across multiple geometries. For 1D flat FCIs, SS size depends inversely on external field $H$ and is critically controlled by the conical reorientation angle relative to the interface normal. This angle determines SS presence, shape (single or multiple), and topological charge sign, with oppositely charged SSs near $\phi_0 = \pi$ exhibiting distinct coupling to inner-twists—one merging with same-charge twists, the other adjacent to opposite-charge twists. In 2D cylindrical FCIs, we observe asymmetric eyebrow-like SSs with depth-dependent profiles, while 2D square-prism FCIs show facet-selective SS formation governed by conical directionality.

Furthermore, our simulations show SSs formation at a skyrmion cluster boundaries within conical phases, where spontaneously formed ferromagnetic shells encapsulate clusters and host adjacent SSs. Our experimental investigations further reveal two distinct formation pathways: thermally activated co-growth with the skyrmion lattice and field-driven transformation from residual helices.

Collectively, the polymorphic nature of SSs enables promising spintronic functionalities. Their distinct states provide the physical basis for a high-density programmable magnetic storage achieving 2-bit-per-cell capacity, while the abrupt topological transitions between SS states - particularly the sharp transition from the SS+ state to the SS- state at a critical threshold - emulate neuronal firing dynamics for neuromorphic computing. These functionalities are governed by the phase factor $\phi_0$, which establishes a fundamental control dimension complementary to traditional approaches, unlocking advanced capabilities including magnon manipulation beyond conventional means. The $\phi_0$-control paradigm thus creates practical opportunities for next-generation spintronic devices with enhanced functionality.

\begin{acknowledgments}
    H.Z. is supported in part by National Natural Science Foundation of China (Grant No. 11704067). 
    H.Z. thanks the computational resources from the Big Data Center of Southeast University and the Center for Fundamental and Interdisciplinary Sciences of Southeast University.
   This work is supported in part by Laboratory Directed Research and Development funds through Ames National Laboratory (H.Z., T.K., L.Z.). L.K. was supported by the U.S. Department of Energy, Office of Science, Office of Basic Energy Sciences, Materials Sciences and Engineering Division, Early Career Research Program following conception and initial work supported by LDRD. Ames National Laboratory is operated for the U.S. Department of Energy by Iowa State University under Contract No. DE-AC02-07CH11358.  All TEM and related work were performed using instruments in the Sensitive Instrument Facility in Ames National Lab. We thank Prof. Sun Weiwei from Southeast University for insightful discussions regarding the potential application of MTJs for detecting surface spirals.
\end{acknowledgments}
\textbf{Data Availability Statement:} 
The data that support the findings of this study are available from the corresponding author upon reasonable request.

\end{document}